%% file: graphs7.tex
\documentclass[12pt]{article}
\usepackage{amsmath,amsfonts,amssymb,a4,color,epsfig,graphics}
\usepackage[latin1]{inputenc}

\newcommand{\R}{{\mathbb{R}}}

\newcommand{\C}{{\mathbb{C}}}
\newcommand{\Z}{{\mathbb{Z}}}
\newcommand{\N}{{\mathbb{N}}}

\def\ha{\frac{1}{2}}

\def\pa{\partial}
\def\ra{\rightarrow}
\def\ga{\alpha}

\def\ge{\varepsilon}

\def\gg{\gamma}

\def\gl{\lambda}
\def\go{\omega}

\def\gs{\sigma}

\newtheorem{defi}{Definition}[section]

\newtheorem{lemm}{Lemma}[section]
\newtheorem{prop}{Proposition}[section]
\newtheorem{rem}{Remark}[section]
\newtheorem{coro}{Corollary}[section]
\newtheorem{theo}{Theorem}[section]

\newenvironment{demo}{\noindent {\it Proof.--}
      \begin{quotation}\noindent}{\end{quotation}\hfill$\square $}
\newenvironment{demolemm}{\noindent {\it Proof of the Lemma.--}
      \begin{quotation}\noindent}{\end{quotation}\hfill$\square $}
\newenvironment{demotheo}{\noindent {\it Proof of the Theorem.--}
      \begin{quotation}\noindent}{\end{quotation}\hfill$\square $}

\begin{document}

\title{Semi-classical measures on Quantum graphs and
the Gau\ss ~ map of the 
determinant manifold}
\author{Yves  Colin de Verdi\`ere\footnote{Université de Grenoble,
Institut Fourier,
 Unit{\'e} mixte
 de recherche CNRS-UJF 5582,
 BP 74, 38402-Saint Martin d'H\`eres Cedex (France);
{\color{blue} {\tt yves.colin-de-verdiere@ujf-grenoble.fr}}}}


\maketitle
\section*{Introduction}
The so-called Quantum Ergodic Theorem (QE), due mainly to Alexander 
Shnirelman \cite{Shn74,Shn93,Zelditch,CV}, 
asserts that the eigenfunctions of the Laplace operator
on a smooth closed Riemannian manifold whose geodesic flow is ergodic
are equi-distributed on the manifold in the limit of large 
eigenvalues provided one removes a  sub-sequence of  density $0$.
The paper \cite{JSS13} gives QE Theorems in the case of metrics with
discontinuities on smooth manifolds: the Assumptions are ergodicity of the geodesic flow 
which in this context is a Markov process and the fact that
recombining geodesics are exceptional. The results apply also to
piecewise smooth metrics on simplicial complexes, in particular to
metric graphs (also called ``quantum graphs''). The geodesic flow is
then ergodic, but they are many recombining geodesics.  
Hence the second  assumption is not satisfied for quantum graphs. It  is
currently 
believed  that QE does not hold in general for  a FIXED  quantum
graph. In \cite{BKW04}, it is proved that QE does not holds in the
limit of large irrational star graphs. This implies that QE does not
hold for a fixed large irrational star graph; this is even proved
for a fixed irrational star graph. Similar results are
proven in  \cite{KMW02}
(put $f(x)=x^2$ in Equation (51)).
In the literature,  QE for graphs means 
 some property of the eigenfunctions in  the limit of a sequence of 
graphs whose number of vertices is going to infinity (see for example \cite{GKF10}). 

In the present paper, our goal is to describe all
weak limits of measures
$|\phi _j|^2 |dx|_{\vec{l}} $, where $(\phi_j )$ is an orthonormal basis
of eigenfunctions for the Laplacian $\Delta _{\vec{l}}$ on a metric
graph $(G,\vec{l})$ and $|dx|_{\vec{l}}$ is the Riemannian measure,
 and to get 
 upper bounds on the densities of the associated sub-sequences
of eigenfunctions.
In particular we show that  QE   does not hold for a generic metric
$\vec{l}$ on a fixed   graph, except
if the graph is homeomorphic to an interval or to a circle.
 In order to avoid confusion, we will denote
this last property by QEF (Quantum Ergodicity for Fixed  graph).

If $G=(V,E)$ is a finite graph, we introduce, following ideas
in \cite{BG00,BG01}, 
  an algebraic  sub-manifold $Z_G$, which we call the
 {\it ``determinant manifold''},
of the torus ${\cal T}_E:=(\R/2\pi \Z)^E$ which allows to compute the eigenvalues
thanks to the so-called {\it ``secular equation''}. 
The  Gau\ss~ map $\Gamma $  associates  to any point of the smooth
part 
  $Z_G^{\rm reg}$ of
$Z_G$  the  half-line of the cone 
 ${\cal P}_E:= [0,+\infty [^E \setminus \{ 0 \} $ which is orthogonal
to $Z_G$. Let us denote
by ${\cal M}_G$ the semi-algebraic set which is 
the closure of the image of $\Gamma $.
Let us state the main result of this paper:
\begin{theo} Let us fix  $G$. For a generic metric  $\vec{l}$,  
 the set of non-normalized   semi-classical measures is the set of all 
$\sum_{e\in E } m_e  |dt_e|$ where $\vec{m}\in {\cal M}_G$ and where
$|dt_e|$ is the Riemannian measure on the edge $e$, i.e.
$t_e$ is an arc-length parametrization of the edge $e$.
Moreover the densities of the corresponding
eigenfunctions sub-sequences are bounded in terms  of the image
by $\Gamma $  of a measure $\mu_{{\rm BG},\vec{l}}$ on $Z_G^{\rm reg}$
introduced in \cite{BG00,BG01}.  
\end{theo}
Another related  result is the description of all semi-classical
measures with {\it minimal support}, which are analogs of 
the so-called {\it ``scars''} in the case
of manifolds: the supports of these scars  are simple paths joining
two vertices of degree one and simple cycles (this is related to the
paper
\cite{SK03}). 
From this, we deduce that QEF does not hold for a generic 
$\vec{l}$,  except for graphs
homeomorphic
to intervals or circles.

{\it \small Many thanks to the referees for the comments which
allowed me to improve the results and the proofs.}

\section{The determinant manifold of  a quantum graph}

In this section, we  recall the way to compute the spectra of the
Laplacians
$\Delta _{\vec{l}}$
on a 
graph $G=(V,E)$ from an algebraic hyper-surface $Z_G$  of the torus
${\cal T}_E= (\R/2\pi \Z)^E $ which depends only on the combinatorics
of the graph and not on the lengths $\vec{l}$.
This approach is closely related to what people do  in several papers like
 \cite{BG00,BKW04,BW08,Ba12,BB13}.

Let us consider a finite connected graph $G=(V,E)$ where loops and multiple
edges are allowed. We choose an orientation of the edges. We
associate, in the usual way,
to $G$ a 1D singular manifold $|G|$ by gluing together $ \# E $
intervals, labelled by the set $E$, using the combinatorics of $G$.
A (Riemannian) metric on $|G|$ is given by  the  lengths  of the
edges. We will denote by $\vec{l} \in ]0,+\infty [^E $
the collection of the lengths and the pair $(G,\vec{l})$ 
is called a {\it Metric  Graph}. 
\begin{defi}
We will say that $\vec{l}$ is {\rm  irrational} if the components
of  $\vec{l}$ are independent over the rational numbers.
\end{defi}
 To  any    metric $\vec{l}$ is associated the
Riemannian 
 measure on $|G|$ denoted by $|dx|_{\vec{l}}=\sum _{e\in E}|dt_e| $
where $t_e \in [-l_e/2, l_e/2 ]$ is the arc-length parametrization of
the edge $e$ following the given orientation and with the origin at
the
center of the edge $e$.  
We have
$\int _{|G|} |dx|_{\vec{l}}=L$ where $L=\sum _{e\in E} l_e $ is the
total length of $(G,\vec{l})$.
The {\it Laplacian} $\Delta _{\vec{l}}$ is the self-adjoint operator
on $L^2(|G|, |dx|_{\vec{l}})$ with domain consisting of 
the functions $\phi:|G|\ra \C$  which
are in the Sobolev space $\oplus_{e\in E} H^2(e)$,  are continuous on
$|G|$
and satisfy the Kirchhoff condition that the sums of outgoing
derivatives of $\phi$ at each vertex vanish. 
Then   $\Delta _{\vec{l}}\phi$  is given by 
$-d^2 \phi/dt_e^2 $ on each edge.
The metric graph $G$ with the previous Laplacian is also called
a {\it Quantum graph}. More generally, a quantum graph is a metric
graph with a self-adjoint differential operator
on $L^2(|G |,|dx|_{\vec{l}})$. 
A good introduction  to Quantum graphs is the recent book
\cite{BK13}.
The operator $\Delta _{\vec{l}}$ has a discrete spectrum given by
$\gl_j =k_j^2 $ with
$k_1=0 < k_2 \leq \cdots $. We denote by $\phi _j :|G| \ra \C $ an
associated orthonormal basis of eigenfunctions.  
Let us remark that we can assume that $G$ has no vertex of degree $2$,
because  we allow multiple edges and loops.

Let us try to compute the non-zero part of the
spectrum of $\Delta _{\vec{l}}$~: 
let us look for an eigenfunction $\phi$
with eigenvalue $k^2$ with $k>0$.
On the edge $e$, we have  
\begin{equation}\label{equ:eigen}  \phi(t_e)= a_e \cos kt_e + b_e \sin k
t_e ~.\end{equation}
 Let us define  $\xi_e =\cos k l_e/2 $
and $\eta_e=\sin k l_e/2 $. From the continuity conditions and the
Kirchhoff
conditions,   we get, for the function $\phi=\phi_{a,b}$,  a linear system
\[ L_{\xi,\eta}(a,b)=0~\]
 of $2 \# E $ equations
with the same number of unknowns, because the values of the function $\phi$
and its derivatives divided by $k$ at the vertices   are
(linear)   functions of
$\xi_e$  and $\eta_e$. Let us remark that if we change any
$(\xi_e,\eta_e)$ into $(-\xi_e,-\eta_e)$, the solutions of the new
system
are changed by $(a_e,b_e)\ra (-a_e,-b_e)$.

 The determinant of the system $L_{\xi,\eta}(a,b)=0$  is an homogeneous  polynomial 
of degree $2$   with
respect
to each $\zeta_e=(\xi_e,\eta_e)$. Hence, it can be expressed as a polynomial of
total degree $\# E$ in the variables  $z_e=(x_e,y_e)$
with $x_e=\cos k l_e,~y_e=\sin kl_e $ which is of degree $1$ w.r. to
each pair $(x_e, y_e)$.
 Let $\theta _e $ be the
corresponding angles so that 
$x_e=\cos \theta _e, ~y_e=\sin \theta _e $. 
Let us denote this trigonometric polynomial
$\delta _G (\theta) $ viewed as a function on the torus ${\cal T}_E =(\R/2\pi
\Z )^E $ whose coordinates are denoted by $\theta =(\theta _e)_{e\in
  E}$.
The following fact is clear from what we said, the number $k^2, ~k>0$
is an eigenvalue of $\Delta _{\vec{l}}$ if and only if
$ \delta _G ([k\vec{l}])=0$ where $[~~]$ means that we take the
components  modulo $2\pi $.
In other words, $[k\vec{l}]$ belongs to the algebraic hyper-surface $Z_G$
of equation 
$\delta _G =0 $ of ${\cal T}_E$.
Let us remark that, by an elementary check,  $\theta =0 $ belongs too to $Z_G$.

Let us discuss a few elementary property of $Z_G$.
We will say that $[k\vec{l}]$ is a {\it smooth point} of $Z_G$
if the differential of $\delta_G $ at that point does not vanish.
A point which is not smooth is called {\it singular}. 
\begin{theo}\label{theo:tangent}
If $k^2>0 $ is a non degenerate eigenvalue of $\Delta _{\vec{l}}$, the
point
$[k\vec{l}]$ is a smooth point of $Z_G$ and the
tangent space to $Z_G$ at that point is given
by $  \sum _e m_e d\theta _e= 0$ where $m_e=a_e^2 +b_e^2$
for an eigenfunction given by Equation (\ref{equ:eigen}).

Conversely,  if the eigenvalue
$k^2 $ of $\Delta _{\vec{l}}$ is  degenerate, $[k\vec{l}]$ is a singular  point of $Z_G$.

The manifold $Z_G$ is singular in a set of co-dimension at least $1$
(w.r. to $Z_G$ itself) except if $|G|$ is a circle.
\end{theo}
Let us denote by 
$Z_G= Z_G^{\rm reg}\cup Z_G^{\rm sing} $ the partition of $Z_G$ into
the sets of regular and singular points.
Using  a result of L. Friedlander on the genericity of non degeneracy
of the eigenvalues of $\Delta _{\vec{l}}$ \cite{Fr05} if
$|G|$ is not a circle, we get the fact
that, if $|G|$ is not a circle,  the set $Z_G^{\rm reg}$  is  dense in $Z_G$ and hence,
 because $Z_G$ is an algebraic
manifold, the set of singular points $Z_G^{\rm sing}$ has co-dimension
 larger than one w.r. to $Z_G$.

{\it Questions:} as we will see from the examples
in Sections \ref{ss:eight} and \ref{ss:cherry},  $Z_G$
can be reducible.
 For which graphs is $Z_G$ irreducible?
A conjecture could be that $Z_G$ is reducible, if and only if $|G |$
admits
 a non trivial symmetry group which is an isometry group
for each metric and $|G|$ is not an interval, more precisely:
\begin{prop} The topological set  $|G|$ admits a non trivial
symmetry group which acts by isometry for each choice of a metric
$\vec{l}$ if and only if either $|G|$ is homeomorphic to $[-1,1]$
(then the isometry group is generated by $x\ra -x$), either $|G-$ is a circle with the antipody as a
non trivial symmetry, 
or if  $|G| $ is obtained by adding  to a $|K| $ (associated to
a graph  $K$) some loops (at least $1$) attached to vertices of $K$
(then the symmetry group is generated by the involutions 
$x\ra -x$ on one of the  loops parametrized by $-1\leq x \leq +1 $).
\end{prop}
\begin{demo} We can assume that $G$ has no vertex of degree $2$.
If $G$ is a circle or an interval it admits clearly such a symmetry
group. 
Let us assume moreover that $G$ has no loops as in the Proposition.
Then  $G$ admits a vertex of degree at least $3$. Because the lengths
of these edge are arbitrary, the isometries have to fix this vertex
and the adjacent adges. An easy induction gives the result.
\end{demo}

 If the manifold $Z_G$ is irreducible, is the set of singular point of
co-dimension $\geq 2$ w.r. to $Z_G$ as expected from the generic co-dimension of symmetric matrices
with som degenerate eigenvalue?

\begin{demo} (of Theorem \ref{theo:tangent})
\begin{itemize} \item 
Using the homogeneity of
the spectrum, i.e. $,\forall c>0,~k_j(c\vec{l})=c^{-1} k_j(\vec{l})$,
we can assume that $k=1$. In other words $Z_G$ is identified with
the $[\vec{l}]$'s so that $1$ is an eigenvalue of $\Delta _{\vec{l}}$. 
Let   $z_0=[\vec{l}]\in  Z _G $. We assume that the eigenvalue  $1$ is
   non-degenerate for $\Delta _{\vec{l}}$.
Close to $z_0$, the set $Z_G$ is defined by the equation 
$k([\vec{l}])=1$ meaning that $1$
is an eigenvalue of $\Delta_{\vec{l}}$,
 so that $Z_G$  is smooth if $dk\ne 0$ and the tangent
space is the kernel of $dk$. 
From the calculation of $dk$ in Appendix \ref{app:deriv}, we get the result.

\item Let us assume that $k_0^2 >0 $ is a degenerate eigenvalue
of $\Delta _{\vec{l}}$. With the notations around Equation
(\ref{equ:eigen}),
this implies that $\dim \ker L_{\xi_0,\eta_0}\geq 2 $
with $\xi_0=(\cos k_0 l_e ),~\eta_0=(\sin k_0 l_e)$.
This implies that the diffenrential of the determinant 
of $L_{\xi_0,\eta_0}$ with respect to any variation of
$\xi,\eta $ vanishes. The result follows.

\end{itemize}

 \end{demo}

The singularities of $Z_G $ are associated to the degenerate eigenvalues.
A description of these singularities would be of interest,
already for general star graphs or complete graphs.
The case of diabolical singularities, corresponding to eigenvalues
of multiplicity two, is used in \cite{BG00} in order to 
study the behavior as $s\ra 0$ of  the distribution $P(s)ds$   of the level
spacing $\gl_{j+1} - \gl_j $ 
defined by
\[ P(s)ds =\lim _{\lambda  \ra \infty }
\frac{ \# \{ j |s \leq \gl_{j+1} - \gl_j \leq s+ds \} }{
 \# \{ j | \gl_j \leq \lambda \} } ~.\]

\section{The Gau\ss ~ map and semi-classical measures}

Let us denote by ${\cal P}_E $ the 
cone $[0,+\infty [^E \setminus \{ 0 \} $.
For  any $ \vec{m}=(m_e)_{e\in E} \in {\cal P}_E $ and any
$\vec{l}$, 
we identify $\vec{m}$ with  the measure $\sum_{e\in E} m_e |dt_e| $ on $|G|$.
In other words, {\bf if $\vec{l}$ is given,}
  we consider ${\cal P}_E $ as a sub-cone of the cone of
positive Radon measures on $|G|$. 
Our goal is to characterize  the measures 
which are semi-classical limits. 
We will need a
\begin{defi}
If $\phi$ is an eigenfunction of some $\Delta _{\vec{l}}$ with eigenvalue
$k^2 >0$, the restriction of $\phi$ to the edge $e$
writes   $\phi(t_e)=a_e \cos k t_e + b_e \sin k t_e$; we denote by
$\mu_\phi$ the measure on $|G|$ defined by 
$\mu_\phi=\sum _e (a_e^2 +b_e^2)|dt_e| $ which belongs to   ${\cal P}_{E}$.
\end{defi}
\begin{defi} A semi-classical limit for $\Delta _{\vec{l}}$
is a weak limit of a sequence of measures
 $\sum _{e\in E }|\phi_{k_j}(t_e)|^2 |dt_e|$ on $|G|$  where $\phi_{k_j}$
are  non-zero  eigenfunctions of $\Delta _{\vec{l}}$ with 
eigenvalues    $k_j^2\ra +
 \infty $.
\end{defi}
\begin{rem} We do not use a lift of the measures to the cotangent
  space 
as in the Shnirelman Theorem. 
\end{rem}
\begin{rem} We do not ask  the eigenfunctions to be $L^2$-normalized.
Our semi-classical limits are in general not probability measures.
\end{rem}

We can summarize the results as follows:
\begin{theo}\label{theo:main}
\begin{enumerate}
\item Let $\mu =\sum_{e\in E }  m_e |dt|_e $  be a semi-classical 
limit for $\Delta _{\vec{l}_0}$, then there exists $\vec{l_\infty}$ and an eigenfunction
$\phi_\infty$ of $\Delta _{\vec{l_\infty}}$ associated to the eigenvalue $1$  so that
$\mu=\mu_{\phi_\infty}$.
\item Let us assume that $\phi$ is an eigenfunction of some
$\Delta _{\vec{l_0}}$ with a simple eigenvalue $k_0^2$, then  
$\mu_\phi $ is a semi-classical measure for $\Delta _{\vec{l_0}}$ and
for any
$\Delta _{\vec{l}}$ with $\vec{l}$ irrational.
\item  Let ${\cal G}_G $ be the set of the $\vec{l}$ so that
$\vec{l}$ is irrational and the line $\{ [k\vec{l}]~|~k>0 \} $ does
not meet $Z_G^{\rm sing}$ (or equivalently the set of $\vec{l}$ so that 
the eigenvalues of $\Delta _{\vec{l}} $ are simple). If $|G|$ is not a circel, 
the set  ${\cal G}_G $ is Baire generic and, if
$\vec{l}_0\in {\cal G}_G $,  the set of semi-classical measures for
$\Delta _{\vec{l}_0}$ 
 is the closure of the image of $Z_G^{\rm
  reg}$ by  the Gau\ss~ map. 
\end{enumerate}
\end{theo}
\begin{demo}
\begin{enumerate}
\item
Let us consider a sequence of   eigenfunctions of $\Delta _{\vec{l}}$
defined by $\phi_{k_j}(t_e)=a_{e,j} \cos k_j t_e +b_{e,j}  \sin k_j t_e $.
Then, for large $k_j$'s, 
the measures $\sum _e \phi_{k_j}(t_e)^2 |dt_e| $ have a weak limit 
if and only if the limits $m_e =\lim _{j\ra \infty}
(a_{e,j}^2 +b_{e,j}^2) $ exist and this limit measure  is given by
$\sum_{e\in E }  m_e |dt|_e $. By compactness of $Z_G$, we can 
assume that $[k_j \vec{l_0}]$ converges to $[\vec{l_\infty}]$ and we
can assume that all components of $\vec{l_\infty}$ are $>0$ .
Then we can again extract a sub-sequence so that the numbers $a_{e,j}$
and $b_{e,j}$ converge to some limit which will be associated, by
the closeness of the eigenfunctions equations, to an eigenfunction of
$\Delta _{\vec{l_\infty}}$ with eigenvalue $1$. 
\item From the assumption and Theorem \ref{theo:tangent}, we get that
the set $Z_G$ is smooth near $[k_0\vec{l_0}]$ and transverse to any
curve $t\ra [t\vec{l}]$ if the components of $\vec{l}$ are all $>0$.
From this and the fact that the linear
flows on tori are recurrent, we can get a sequence
$k_j\ra +\infty $ so that $[k_j \vec{l_0}] \ra [k_0\vec{l_0}]$ as
$j\ra \infty $ and we have that $\mu _{\phi_{k_j}} \ra \mu _{\phi}$.
Similarly, if $\vec{l}$ is irrational, we can find $k_j\ra \infty $ so that
$[k_j \vec{l}]\in Z_G$ and $[k_j \vec{l}]\ra [k_0 \vec{l_0}]$.

\item The genericity comes from the fact that $Z_G^{\rm sing}$ is of
  co-dimension at least two in ${\cal T}_E$. From the previous part, we know already
  that each measure in the image of the Gau\ss ~map is a semi-classical 
measure and the same is true for  a measure in the closure of this image, because the set of
semi-classical
measures is closed.
On the other hand, we know that each semi-classical measure is the
limit
of a sequence of $\mu_{\phi_{k_j}}$; because of the assumption on
$\vec{l}$ this implies that each semi-classical measure lies in the
closure of the image of the Gau\ss~ map. 

\end{enumerate}
\end{demo}

\section{The Barra-Gaspard measure, Weyl formula
 and the densities
of semi-classical limits}
In \cite{BG00}, the authors introduce a $\vec{l}-$dependent 
measure on $Z_G^{\rm reg}$, which we will denote by
$\mu _{{\rm BG},\vec{l}}$.
The manifold $Z_G^{\rm reg}$ is transverse to any vector with strictly positive
coordinates and hence oriented. The measure 
 $\mu _{{\rm BG},\vec{l}}$ is defined
\footnote{Avoiding the  intrinsic calculus
of exterior differential forms, the Barra-Gaspard measure can be defined
in the following  less ``cryptic'' way: 
if $d\sigma _Z $ is the measure on
$Z_G^{\rm reg}$  associated to  the metric induced by the
flat
metric $\sum _{e\in E}d\theta _e^2$ on ${\cal T}_E$ and $\pm \vec{\nu} $ are  the unit normal vector
fields  to $Z_G^{\rm reg}$, we put
\[ \mu _{{\rm BG},\vec{l}}:=(2\pi)^{-\# E}|\vec{\nu}.\vec{l}| d\sigma _Z~.\]
}
 by
\[ \mu _{{\rm BG},\vec{l}}=(2\pi)^{-\# E}| \iota (\vec{l})\wedge _{e\in E} d\theta _e
|~\]
($\iota (\vec{V})\go $ is the inner product),
which is a Radon measure with strictly positive density everywhere on $Z_G^{\rm
  reg}$.
\begin{prop}\label{prop:mass}
 The mass of the measure $\mu_{{\rm BG},\vec{l}}$
is exactly $L/\pi $.
\end{prop}
\begin{demo}
The total mass of the  measure $\mu_{{\rm BG},\vec{l}}$ is 
less than  $L/\pi $:
let us consider, for $e\in E$,  the canonical projection
 $\pi _e: Z_G \ra (\R /2\pi \Z)^{E\setminus
  e}$. 
From the definition of $\mu _{{\rm BG},\vec{l}}$,  we get
\[ \int _{Z_G} \mu _{{\rm BG},\vec{l}}=\sum_{e\in E} l_e \int 
_{ (\R /2\pi \Z)^{E\setminus e}}{\rm deg}_e (\theta ) |\wedge \widehat{d\theta
    _e} |\] 
where ${\rm deg}_e (\theta )$ is the cardinal of $\pi _e ^{-1}(\theta )$.
The fact that $\delta _G $ is of degree $1$ w.r. to each $(\cos
\theta_e,\sin \theta_e)$ 
implies that ${\rm deg}_e \leq 2 $ and the upper bound of the
mass.

In fact ${\rm deg}_e \equiv 2$ as follows from Lemma \ref{lemm:as-ZG}
below and the
Weyl asymptotic formula (see \cite{BK13} p. 95)
\begin{theo}
We have (Weyl's law)
\[\lim_{K\ra \infty}\frac{1}{K} \# \{ 0\leq  k\leq K | ~[k\vec{l}] \in Z_G \} =
\frac{L}{ \pi }  ~.\]
\end{theo}
\end{demo}

We will need the following Lemma, similar to Proposition 4.4 of
\cite{BW08} and coming also from the ideas of Barra-Gaspard:
\begin{lemm}\label{lemm:smooth}  If $\vec{l}$ is irrational and $D \subset Z_G^{\rm reg}$
is a compact domain with piecewise smooth boundary, we have
\[ \lim _{K\ra \infty } \frac{1}{K} \# \{ k_j ~|~0< k_j \leq
K,~[k_j\vec{l}] \in D \} =   \int _D \mu _{{\rm BG},\vec{l}}
~.\]
Similarly, if $f:Z_G
\ra \R $ has a  compact support in $Z_G^{\rm reg}$, we have:
\[ \lim _{K\ra \infty } \frac{1}{K} \sum _{0< k_j \leq
K,~[k_j\vec{l}] \in Z_G }f([k_j\vec{l}]) =   \int _{Z_G} f \mu _{{\rm BG},\vec{l}}
~.\]
\end{lemm}
\begin{demo} Let us choose $\ge >0 $ small enough so that 
the map $F:[-\ge ,\ge ]\times D \ra {\cal T}_E $,
defined by $F(t,\theta )=\theta +t \vec{l}$, 
is a smooth embedding of image $D_\ge $. 
From the unique ergodicity of the Kronecker  flows on the tori, we get
\begin{equation} \label{equ:lim} \lim _{K \ra \infty} \frac{1}{K}
| \{ k~|~0<k\leq K , [k\vec{l}]\in D_\ge \} |=(2\pi)^{-\# E}{\rm vol
}(D_\ge ) \end{equation}
where $|X|$ denote the Lebesgue measure of a set  $X\subset \R$.
We observe that $[k\vec{l}] $ belongs to $D_\ge$ if and only if 
there exists $j$ so that $[k_j\vec{l}]\in D$ and
$k\vec{l}=[k_j\vec{l}] +t \vec{l}$ with $|t|\leq \ge $.
 The lefthandside of Equation (\ref{equ:lim}) is equal to
\[  \lim _{K \ra \infty} \frac{2\ge }{K}\# \{ k_j ~|~0< k_j \leq
K,~[k_j\vec{l}] \in D \} \] 
while, using the definition of $\mu_{{\rm BG},\vec{l}}$, the
righthandside
of Equation (\ref{equ:lim}) is 
equal to 
$ {2\ge }  \mu _{{\rm BG},\vec{l}}(D) $.
\end{demo}

We can remove the Assumption that $D$ is compactly
supported in $Z_G^{\rm reg}$ as follows:
\begin{lemm}\label{lemm:as-ZG} If $\vec{l}$ belongs to ${\cal G}_G$,  we have
\[ \lim _{K\ra \infty } \frac{1}{K} \# \{ k_j ~|~0< k_j \leq
K,~[k_j\vec{l}] \in Z_G \} =   \int _{Z_G} \mu _{{\rm BG},\vec{l}}
~.\]
A similar result holds for the integration of a continuous function
$f$ on $Z_G$:
\[ \lim _{K\ra \infty } \frac{1}{K} \sum  _{ 0< k_j \leq
K,~[k_j\vec{l}] \in Z_G  }f([k_j \vec{l}]) =   \int _{Z_G}f \mu _{{\rm BG},\vec{l}}
 ~.\]
\end{lemm}
\begin{demo}
Let $R_\ge $ be the ball of radius $\ge $ in $Z_G$ centered
on $Z_G^{\rm sing}$ and
$\tilde{R}_\ge $ be the set 
\[ \tilde{R}_\ge :=\{ z+ t\vec{l}|z\in R_\ge ,~0\leq t \leq \ge \}~.\]
 We have ${\rm vol }(\tilde{R}_\ge)= O(\ge ^2)$, because the
 codimension
of $Z_G^{\rm sing }$ in ${\cal T}_E $ is larger than $2$.
 Using the result of Appendix \ref{App:quasi}, 
 there exists a constant $C >0$, so that, for any $n \in \N$, 
we have:
\[ \# \{ k |n<k \leq n+1,~ [k\vec{l}]\in R_\ge \}
\leq \frac{C}{\ge }|\{ k' | n-\ge\leq k' \leq n+1+\ge , 
 [k'\vec{l}]\in \tilde{R}_\ge \}| ~\]
(if $[k'\vec{l}]$ belongs to $\tilde{R}_\ge$, there exists
$k$ so that $[k\vec{l}]\in R_\ge $ and 
$|k-k'|\leq \ge $; the constant $C$ is an uniform  upper  bound on the number
of solutions of $\delta _G (k\vec{l})=0 $ with $n<k \leq n+1$).
Assuming that $K$ is an integer and
summing the previous inequalities from $n=0 $ to $n=K-1 $,
we get a constant $\tilde{C}$ so that 
\[ \frac{1}{K}\# \{ k |0<k \leq K,~ [k\vec{l}]\in R_\ge \}
\leq \frac{\tilde{C}}{\ge K}|\{ k' | 0\leq k' \leq K+1 , 
 [k'\vec{l}]\in \tilde{R}_\ge \}| ~.\]
The limit of the righthandside as $K\ra \infty $ is ${\tilde{C}}/{\ge
}$
times the
volume of $\tilde{R}_\ge $  and can 
be close to $0$ by choosing $\ge $ small enough.
The proof is completed by writing $Z_G=(Z_G\cap R_\ge )\cup
(Z_G\setminus Z_G\cap R_\ge )$, by using Lemma \ref{lemm:smooth} for
the main term and by using the finitness of the mass of $\mu_{{\rm
    BG},
\vec{l}}$ (first part of the proof of Proposition \ref{prop:mass}).
\end{demo}

The Barra-Gaspard measure is related to the densities of sequences of
eigenvalues
giving a given semi-classical limit. This density vanishes  in general,
so we have to say it in the following way:
\begin{theo} Let $D\subset {\cal P}_E$ be a compact domain with
  smooth boundary and consider a sequence of eigenfunctions $\phi_{k_j}$
of some $\Delta _{\vec{l}}$ with $\vec{l}$ generic. Let us assume that
all semi-classical limits of the $\mu_{\phi_{k_j}}$ lie in $D$, then
the density of the sequence $k_j$ is bounded from above in terms of
 $\mu_{{\rm BG},\vec{l}}$ as follows:
\[ \limsup _{K\ra \infty } 
\frac{\# \{k_j\leq K\}}
{K} \leq  \mu_{{\rm BG},\vec{l}}\left( \Gamma^{-1}(D)\right) ~.\]

The converse is true; for any generic $\vec{l}$ and for any  compact domain $D\subset {\cal P}_E$,
there exists a sub-sequence  $\phi_{k_j}$ of density
$\mu_{{\rm BG},\vec{l}}\left( \Gamma^{-1}(D)\right)$  so that the
semi-classical limits associated to sub-sequences of  $\phi_{k_j}$
belongs to $D$.
\end{theo}
\begin{demo}
Let $D'\subset  {\cal P}_E$  a neighborhood of $D$. Then for
$j$ large enough, $\Gamma_{\vec{l}}([k_j \vec{l}])$ belongs
to $D'$. So that, from a slight extension of Lemma \ref{lemm:as-ZG}
to domains in $Z_G$, we get the upper bound
$\mu_{{\rm BG},\vec{l}}(\Gamma_{\vec{l}}^{-1}(D')$.

For, the converse, it is enough to take the sequence of  $k_j$'s so that
$[k_j\vec{l}]$ belongs to $\Gamma ^{-1} (D)$.
\end{demo}

We get the following
\begin{coro}
If $\vec{l}$ belongs to $Z_G^{\rm reg}$ and $\vec{l} $ is generic,
let $\phi $ be an eigenfunction of $\Delta _{\vec{l}} $ with
eigenvalue $1$. Then, for all neighborhoods $U$ of $\mu_\phi $,
there exists sub-sequences of positive densities  of the
eigenfunctions of $\Delta _{\vec{l}}$ so that 
the semi-classical  limits of sub-sequences belongs to $\Gamma (U)$.
\end{coro}
\begin{coro}\label{coro:nonQE} If $\vec{l}$ is generic
and there exists an eigenfunction $\phi $ of $\Delta_{\vec{l}}$
so that $\mu_\phi $ is not the Liouville measure, then QEF does not
holds for $(G,\vec{l})$.
\end{coro}

Moreover, we have the following link with the Liouville measure:
\begin{theo} \label{theo:liouville} Let 
$\Gamma _{\vec{l}} $ be the map $\Gamma $ normalized so that
the sums $\sum m_e |dt_e| $ are probabilities, then
\[ \int \Gamma _{\vec{l}} (\theta ) d\mu _{{\rm
    BG},\vec{l}}=\frac{1}{\pi}\left( \sum_e 
  |dt_e |\right)~. \]
This can be reformulated as
``The average of the Gau\ss~ map  with respect to $\mu_{{\rm BG},\vec{l}}$
is the Liouville measure''.

\end{theo} 
\begin{demo}
Le us define 
\[ A_e(K)=\frac{l_e}{2K} \sum _{0\leq  k_j \leq K}(a_{j,e}^2
+b_{j,e}^2 )\]
where the normalized eigenfunction $\phi_j$
writes, on the edge $e$,
$\phi_j(t_e)=a_{j,e}\cos kt_e + b_{j,e} \sin kt_e $.
From the local Weyl Theorem and the fact that
\[ \int _e |\phi _j (t_e)|^2 dt_e =\frac{l_e}{2}(a_{j,e}^2
+b_{j,e}^2 )+ O(1/k)~,\] 
we get that
\[ \lim _{K \ra \infty} A_e(K)= l_e/\pi~.\] 
On the other hand, we have
\[ \Gamma _{\vec{l}}([k_j \vec{l}])= \ha (a_{j,e}^2
+b_{j,e}^2 )+ O(1/k)~.\]
Using Lemma \ref{lemm:as-ZG}, we get
\[ \lim_{K\ra \infty} A_e (K)=l_e \int _{Z_G} \left( \Gamma
  _{\vec{l}}(\theta)\right) _e d\mu _{{\rm BG},\vec{l}} ~.\]
\end{demo}

\section{Scars}

Let us look at semi-classical measures with small supports; we have
the: 
\begin{theo}\label{theo:scars}
 Let $G$ be given.  The minimal supports of the semi-classical
 measures for a generic 
  $\vec{l} $  are the simple cycles and the simple paths between two  vertices of
degree $1$ of $G$.  These measures are extremal points of the convex hull of
all 
semi-classical measures.
\end{theo}
We start with the 
\begin{lemm}\label{lemm:supp}  Let  $K \subset |G|$ be the support of a semi-classical
  measure $\mu $. Then $K$ is an union of edges of $G$
and  every vertex of $K$ is of degree $\geq 2 $ in $K$
or is of degree $1$ in $G$.
\end{lemm}
\begin{demolemm} From the first Assertion  of Theorem \ref{theo:main}, 
we deduce that $\mu = \mu_\phi $ where $\phi $ is an eigenfunction of
some
$\Delta _{\vec{l_\infty}}$ on $G$. The Lemma follows then from the
Kirchoff
conditions applied to $\phi $.
\end{demolemm}

\begin{demotheo}

\begin{itemize}
\item {\it Simple paths joining to vertices of degree $1$ are minimal
    supports:}
Let  $\gg $ be a  simple oriented path   whose $N$ vertices are
$(1,2,\cdots, N)$ and choose  a vector $\vec{l}$ so
that
the lengths of the edges of the path  are $\vec{l}_0=(1/2, 1, 1,\cdots, 1,1/2)$.
If we parametrize this path by $0\leq t \leq N-2$, the function
$\phi_1$ defined by $\phi_1=\cos t $ on $\gg $ and $0$ outside is an
eigenfunction
of $\Delta _{\vec{l}}$ with eigenvalue $\pi ^2 $.
I claim that we can choose such a vector $\vec{l}$ extending
$\vec{l}_0$
 so that the eigenvalue $\pi^2$ of $\Delta _{\vec{l}}$
 is simple: if it is not, let $X:=\ker \left(\Delta
  _{\vec{l}}-\pi^2 \right)$, and $l_t $ defined by 
$(l_t)_e=(1-t)l_e$ for $e\notin \gg $ and $(l_t)_e=l_e$ for $e\in \gg
$.
Then the quadratic form on $X$ defined by $f \ra \langle \dot{\Delta
}f |f \rangle $ vanishes on $\phi_1$ and is $>0 $ in any other
direction (because as we have  seen the support of $\phi_1$ is minimal).
It implies that $\pi^2$ is a simple eigenvalue for $\Delta
_{\vec{l}_t}$ for small non-zero $t$.
 It follows that the uniform
measure on $\gg $ is a semi-classical measure for this choice of
$\vec{l}$  and hence for all generic  $\vec{l}$'s. 

The minimality follows from Lemma \ref{lemm:supp}.
\item {\it Simple cycles are minimal supports:}
Let  $\gg $ be a  simple oriented cycle   whose $N$ vertices are
$(1,2,\cdots, N-1, N=1)$ and choose  a vector $\vec{l}$ so
that
the lengths of the edges of $\gg$   are $1$.
By an argument similar to the case of a simple path, we get 
that the uniform
measure on $\gg $ is a semi-classical measure for this choice of
$\vec{l}$ and
all irrational $\vec{l}$'s.

Again the minimality follows from  lemma \ref{lemm:supp}.
\item {\it If $K$ is the support of a semi-classical measure, $K$
contains a simple path joining 2 vertices of degree $1$ of $G$ or a
simple
cycle of $G$: }
if $K$ contains no simple cycles, $K$ is a forest and every
vertex of degree $1$ of $K$ is also of degree $1$ in $G$. So that
every sub-tree of $K$ contains simple paths joining 2 vertices of
degree $1$ in $G$. 
\end{itemize}

\end{demotheo}

 As a Corollary, we get
\begin{theo}
For a generic $\vec{l}$,  QEF holds for 
$\Delta _{\vec{l}}$  if and only if
$|G|$ is homeomorphic to a circle or to an interval.
\end{theo}
Any connected graph which is not homeomorphic to a circle or an
interval
contains as strict sub-graphs  a simple cycle or a path whose ends are
of degree one.
The Theorem \ref{theo:liouville}  implies that the image of $Z_G^{\rm reg}$  by the Gau\ss~ map
contains at least one line distinct from the Liouville measure.
The proof is completed by applying  Corollary \ref{coro:nonQE}.

\section{Ergodicity of the geodesic flow on graphs}
In the case of a connected quantum graph, the classical dynamics is
ergodic
as shown from the study of the associated Perron-Frobenius operator
done in \cite{BG01}.
The phase space $Z$  of a Quantum graph $G
 =(V,E,\vec{l})$  (the unit cotangent bundle) can be identified
with the set of oriented edges: to a point $x$ of an oriented edge, 
we associate the unit co-vector pointing in the direction 
given from the orientation.
Using the probabilities given  from the transition coefficients
in Appendix \ref{app:transitions}, we get the geodesic flow  on $Z$
as a Markov process. 
We have the
\begin{theo}
If $G $ is connected and is not homeomorphic to
 a circle, the geodesic flow on
$(V,E,\vec{l})$ is ergodic.
\end{theo}
In order to prove the previous result, we introduce the
Perron-Frobenius semi-group $T_t,~t\geq 0 $ acting on 
$L^\infty (|G|,|dx|_{\vec{l}})$
 as follows:
\[ T_tf(x)=\sum _{\gg:[0,t]\ra |G|,~\gg(0)=x} 
w(\gg ) f(\gg(t)) ~,\]
where the sum is over the continuous paths  with speed $1$ 
starting from $x$ at time $0$ and $w(\gg)$ is the product of the 
transitions probabilities at all $t$ for which $\gg(t)$ is a vertex.

\begin{demo} It is enough to prove that the only functions
invariant by the Perron-Frobenius semi-group  $(T_t)_{t>0}$
are the constant functions.
We see first that such a function $f$ has to be constant on each
oriented edge by using at a point $x$ the invariance by $T_t$ for $t$
small.
Moreover if $G $ is connected and is not homeomorphic to
 a circle, the geodesic flow is transitive: for any pair  $x,y \in Z$
there exists a geodesic from $x$ to $y$.
Let us choose $f $ constant on the oriented edges and invariant 
by the semi-group  $(T_t)_{t>0}$. Let us choose an edge $e_0$ sot that
$\forall x\in Z,~f(x)\leq f(e_0)$ and $x_0 \in e_0$.
We have $T_t f (x_0)=f(x_0)$ and $T_tf(x_0)=
\sum _{\gg :[0,t]\ra Z } w(\gg) f(\gg(t))$. Using the fact that
$ \sum _{\gg :[0,t]\ra Z } w(\gg)=1$, we get that $
f(y)=f(x_0)$ if there is 
a   geodesic $\gg$ so that $\gg (0)=x_0 $ and $\gg (t)=y$. This holds
for all $y \in Z$ if the geodesic flow is transitive.

\end{demo}

From the previous study, we see that the QEF Theorem is not valid
for Quantum Graphs even if the geodesic flow is ergodic, which
is the case  for all connected graphs which are not homeomorphic to
 a circle. In view of the main result of \cite{JSS13} which says that QE
holds if the geodesic flow is ergodic and  the set of recombining
geodesics
 is of measure $0$,
 this is due to the fact that there are a lot of recombining
geodesics: for example in a star graph, 
they are many  geodesics starting from $x$ in an edge $e_1$
 an coming back to $x$ in
the
opposite direction  following a geodesic 
containing  $k_1$ times the edge $e_1$
and  $k_2$ times the edge $e_2$ with $k_1,k_2\geq 1$ and 
with different orders.

\section{Examples}

\begin{figure}[hbtp]
\leavevmode \center
\input{examples.pstex_t}
\caption{{ The 3 examples: the black circles are the
    vertices
and the white circles are the middle of the edges}}
 \label{fig:examples}
\end{figure}
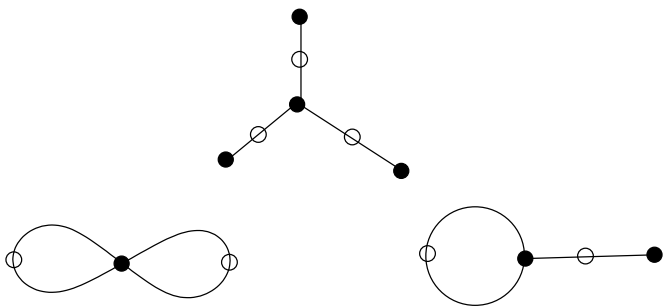

\subsection{The star graph with three  edges}
Let us consider for simplicity the case of a  star graph with three  edges
$G =(V,E) $ where 
$l_1,l_2,l_3$ are    the  lengths of   
three  edges.
The function $\delta _G $ can be computed as follows:
we assume that, we parametrize the $j$th   edge by $0\leq t_j \leq l_j$
where $0$ corresponds to the vertex  of degree $1$.
Then an eigenfunction of eigenvalue $k^2$ will be of the 
form $\phi(t_j)=a_j \cos kt_j $. Putting $x_j = \cos kt_j $
and $y_j =\sin kt_j $, we get the three  equations
\[ a_1 x_1=a_2 x_2 =a_3 x_3, a_1 y_1 + a_2 y_2 + a_3 y_3 =0~.\]
Hence
\[ \delta _G (x,y)= ~x_1 x_2 y_3+ x_2 x_3 y_1 + x_3 x_1 y_2~.\]

{\it Singularities of $\Sigma$:}
The surface $\Sigma_G  \subset {\cal T}_E  $ is invariant by the eight
translations whose vectors have coordinates $0$ or $\pi $.
We will restrict ourselves to the intersection of $\Sigma _G  $
with the cube $C=[-\pi/2,\pi/2 ]^3$. The seven other cubes are deduced by 
translations. The surface $\Sigma _G $ is also invariant by the
central symmetry $\gs : (\theta _j) \ra (-\theta _j )$.

We can cover $\Sigma $ by four  types of charts and use adapted
coordinates taken from  the  set $X_j =x_j/y_j, Y_j =y_j/x_j $:
\begin{enumerate}
\item In the domain $x_1 x_2 x_3 \ne 0 $, we get 
$Y_1 +Y_2 +Y_3  =0$ which is  smooth. 
\item In the domain $y_1 y_2 y_3\ne 0$, 
we get $X_1 X_2 + X_2 X_3 + X_3 X_1 =0$
which is singular
at the eight points where  all $x_j$ vanish. They are ``diabolical''
points
located at the vertices of the cubes. 
\item In the domains, $x_1x_2 y_3\ne 0$ and $x_1 y_2 y_3 \ne 0$,
$\Sigma _G$ is smooth. 
\end{enumerate}

The boundary of the 
closure of the intersection $\Sigma_0= \Sigma \cap ]-\pi/2,\pi/2  [^3 $
is the union of  the six edges of $C$ which have no
vertex $(-\pi/2,-\pi/2,-\pi/2 )$ or $(\pi/2, \pi/2 ,\pi/2)$.
We call this closed path the {\it equator} of $C$.
The surface $\Sigma _0 $ is a smooth $2$-disk whose boundary 
is the equator of $C$.
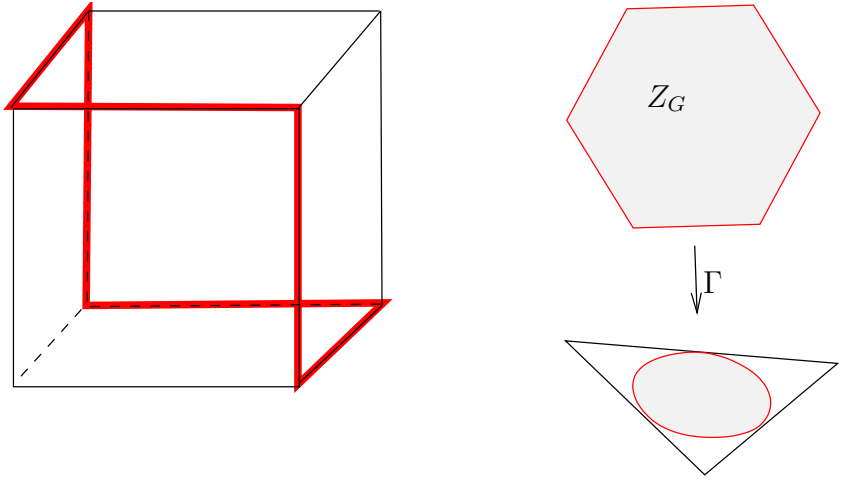
\begin{figure}[hbtp]
\leavevmode \center
\input{Sigma.pstex_t}
\caption{{\it \small the surface $Z_{\rm star}$ and the equator of the cube;
the map $\Gamma $ from $Z_{\rm star}$ to the semi-classical measures.  }}
 \label{fig:Sigma}
\end{figure}

The set ${\cal M}_G $  is the 
cone 
$\sum _{i=1}^{3} m_i^2 \leq 2\sum _{1\leq i < j \leq 3}m_i m_j $.
In particular, this closed set is not equal to ${\cal P}_E$.
 
\subsection{The eight figure} \label{ss:eight}

Let us denote by $G _8 $ the graph with one vertex and two
edges (two loops). We have 
\[ \delta _{G _8} (\theta_1, \theta _2)=
\sin \frac{\theta _1}{2} \sin \frac{\theta _2}{2}
\sin \frac{\theta_1 +\theta _2}{2}~.\]
The set $Z_{G _8 } $ is the union of the circles
$\theta _1=0$, ~$\theta_2=0$ and $\theta_1+\theta _2 =0 $.
The image by the Gau\ss ~ map is the union of the three lines
$m_2=0$, $m_1=0$ and $m_1=m_2$. The associated densities
are up to global normalization
$l_1,~l_2,~l_1+l_2$. This implies that there exactly three
semi-classical limits for an irrational length vector.
The Liouville measure $|dt_1|+|dt_2|$ has relative density $\ha $. 
On this example we see that the set of semi-classical measure is not
even connected and that the scars give measures which are extremal
point of their convex hull.

\subsection{The cherry} \label{ss:cherry}

Let $G _C$ be the cherry graph consisting of one loop with an
attached edge. Let us denote by $l_1$ the length of the loop and by
$l_2$ the length of the edge.
We get for $Z_{G_C}$ the equation:
\[ ~\sin \theta _1 \sin \theta _2 +2 (1-\cos \theta_1)\cos \theta _2
=0.\]
The set  $Z_{G_C}$ is the union of the three circles 
 \[ \{ \theta _1=0 \} \cup \{ \tan \theta _2 +2 \tan \theta _1/2 =0 \}
 ~\]
The circle $\{ \theta _1=0 \}$ correspond to the scars localized on
the loop~; the other part corresponds to a continuum of semi-classical
measures $m_1 |dt_1 |+ m_2 |dt _2|$ with
$m_1l_1 +m_2  l_2=1$, $1 \leq m_2/m_1 \leq 4$.

\section{Limits of graphs and  a new proof of Friedlander's Theorem}
 
From the characterization of the spectrum of $\Delta_{\vec{l}}$ as the
set of $k^2$ so that $[k\vec{l}]$ belongs to $Z_G$, on can still define
the spectrum of $\Delta _{\vec{l}}$ for $\vec{l}$ a real vector.
In particular, one can take some lengths to be zero and the others
$>0$.
What is then the interpretation of this ``spectrum''?

Let $G $ be a graph as before. Given a set of edges $X\subset E$,
we introduce a new graph $G _X$ obtained by contracting all edges
$e\in X$. From the point of view of Riemannian 
metrics, the graph $G _X $ can be interpreted as the graph
$G $ where  the lengths of the edges in $X$ vanish.
We have the
\begin{lemm}
For any connected graph $G =(V,E) $  not homeomorphic to the circle or
the interval,
there exists a set  $X\subset E$ so that $G _X$ is 
homeomorphic to the star graph with three edges, to  $G _8 $ or to
$G _C$.
\end{lemm}
Let us denote by $b_1(G)$ the first Betti number of $|G|$, i.e. the 
dimension of the space of cycles of $|G|$.
 
The first case occur if $G $ is a tree not reduced to an interval,
 the second one if
$b_1(G)\geq 2$ and the third one if $b_1(G)=1$ and $|G| $ is not a
circle.

Let $(Z_{G})_X $ be the intersection of $Z_G $ with the
torus
${\cal T}_X=\{ \theta _e =0,~ \forall e\in X \} $.
We have clearly an identification of  $(Z_{G})_X $
with $Z_{G _X}$. 

Any non singular point of $Z_{G _X}$  is in this way associated
to a non singular point of $Z_G $. 
This gives an independent proof of Friedlander's result by using the
fact
that, for  each of the previous reduced graphs, the manifold $Z_G$ 
admits non singular points.

\appendix

\section{The derivatives of the eigenvalues w.r. to the edge lengths}
\label{app:deriv}

This Lemma is contained in \cite{Fr05}:
\begin{lemm}
If $\gl $ is a non degenerate eigenvalue of $\Delta _{\vec{l}}$
with a normalized eigenfunction $\phi(t_e)=a_e \cos kt_e + b_e \sin kt_e$
 on the edge $e$, 
the derivative of $\gl $ w.r. to $l_e$ is  given
by
\[ \frac{\pa \gl  }{\pa l_e} =- \gl  \left(a_e^2 + b_e^2\right) ~.\]
\end{lemm}
\begin{demo} Let us put $t=t_e$, $I=]-l_e/2, l_e/2 [$   and 
 choose a function $\psi \in C_o^\infty (I,\R) $
and the metric
$g_u = {\rm exp} (4u \psi(t) ) |dt|^2 $
on the edge $e$ and independent of $u$ on the other edges. 
Then we have
$d l_e/du = 2 \int _I \psi (t) |dt| $,
\[ \| \phi \|_{u}^2 = {\| \phi \|'}_0^2 + \int _I e^{2u\psi(t)}\phi^2(t) |dt|
~,\]
and the Dirichlet integral
\[ q_u (\phi)=q'_0 (\phi) + \int _I e^{-2u\psi(t) }{\phi'}^2(t) |dt| ~,\]
where $'_0$ denotes the integrals on the other edges which are
independent of $u$.
Let now take the $u-$derivative of the Rayleigh quotient
$q_u(\phi)/\| \phi \|_{u}^2$ 
at $u=0$, we get 
\[ d\gl /du= -2 \int _I \psi(t) ({\phi'(t)}^2 +\gl \phi(t)^2)|dt| ~\]
from which the result follows. 
\end{demo}

\section{Appendix: the transition probabilities
for a quantum graph}\label{app:transitions}

We want to describe the way a wave arriving at a vertex of a
quantum graph splits into several waves. 
Let us denote by $O$ the vertex of degree $d$  and by $e_j,~j=1,\cdots ,d $
the $d$-edges arriving at $O$. Let us denote  the arc-length 
coordinate $x_j$  along $e_j$ starting from $0$ at $O$.
Let us consider a function $f_1$ compactly supported 
in $]0,+\infty [$ 
 and a wave
$u(x,t)$ defined on $e_1$ by
$u(x_1,t)=f_1(x_1 +t) + g_1(x_1-t)$
and,  on $e_j$ for $2 \leq j\leq d $, by $u(x_j,t)=g_2 (x_j-t)$.
From the Kirchhoff conditions, we get
$ f_1(t) + g_1(-t)=g_2(-t)$
and
$ f'_1(t)+g'_1 (-t)+ (d-1) g'_2 (-t)=0$.
Integrating the second equation, we get
$f_1(t)=g_1(-t)+ (d-1)g_2 (-t)$
(we see that the integration constant vanishes by putting $t=0$).
So that we get
\[ g_1(t)= \frac{2-d}{d} f_1(-t),~ g_2(t)=\frac{2}{d}f_1 (-t)~.\]
The transition probabilities are the
numbers
\[ p_{j,j}=\left( \frac{d-2}{d} \right)^2 \]
and for $i\ne j$, 
\[ p_{i,j}=\left( \frac{2}{d} \right)^2~.  \]

The quantum graph is said {\it classically ergodic} if the
following Markov process defined on the unit tangent bundle of the
metric graph  is ergodic: follow the  edges with unit speed and arriving
at a vertex use the  transition probabilities defined before.
We can assume that the graph has no vertex of degree two.

\section{Appendix: the case $\vec{l}=(1,1,\cdots, 1)$}

In this case the spectrum is given by the $k_j^2, ~k_j\geq 0 $ so 
that $[k] \in Z_G ~.$
Using Weyl asymptotics, the number 
of $k_j$'s with multiplicity is $2\# E $.
On the other hand, we find from \cite{Ni84,Ni87,Catt97}, that
the eigenvalues of $\Delta _{\vec{1}} $
are given in terms of the eigenvalues $\mu_l,~l=1,\cdots, \# V,$
of the weighted adjacency operator defined
\[ A_G f(i)=\frac{1}{d_i}\sum_{j\sim i} f(j)~.\]
The result is the following one for the $k_j$:
\begin{itemize}
\item For each $|\mu _l |< 1$, 2 values of $k$
defined by $\cos k=\mu_l$.
\item The value $k=0$
with multiplicity
$1+ b_1(G)$.
\item The value $k=\pi $ with multiplicity
$b_1(G)+1$ 
if $G$ is bipartite
and $b_1(G) -1$ if-not.
\end{itemize}
At the end, we have, in the bipartite case
$ 2\# E= 2 (\# V -2) + 2b_1(G) +2$,
and in the non-bipartite case
$ 2\# E= 2 (\# V -1) + 2b_1(G) $.
Both formulas give the Euler formula for $G$:
\[ \# V -\# E =1-b_1(G) ~.\]

\section{Appendix: a Lemma on polynomial quasi-periodic
functions} \label{App:quasi}

Let $P$ be a non zero real valued polynomial of $2N $ variables
and consider, for $\ga \in R^N $, the function
\[ f(t):=P(\cos t \ga_1,\sin t \ga_1,\cos t\ga _2,\cdots )~.\]
We have the
\begin{lemm}
The number  $N(x)$  of zeroes (with multiplicity) of $f$ in the
interval $[x,x+1]$ is uniformly bounded for $x\in\R $.
\end{lemm}
\begin{demo}
The function $f(t)$ extends to a bounded holomorphic function
in the strip $-2 < \Im t <2$. The set of functions
$f_x (t)=f(t-x)/\| f \|_{L^\infty( [x_1,x+2]\times [-1,+1])}$  is therefore a compact set of holomophic
functions
on $D=[-1,2]\times [-1,+1]$. 
Let us assume that $N(x)$ is unbounded, we can choose a sequence
$x_j$ so that $N(x_j)\ra \infty $ and $f_{x_j}$ converges
on $D$ to a non-zero holomorphic function $f_\infty $.
For $j$ large enough, the
 number of zeroes $N(x_j)$  of $f_{x_j}$ on $[0,1]$
 is less than  the the (finite) number
of zeroes of $f_\infty $ on $[-1,2]$. The contradiction follows.
\end{demo}

\bibliographystyle{plain}

\end{document}

%% file: examples.pstex_t
\begin{picture}(0,0)%
\includegraphics{examples.pstex}%
\end{picture}%
\setlength{\unitlength}{4144sp}%
\begingroup\makeatletter\ifx\SetFigFont\undefined%
\gdef\SetFigFont#1#2#3#4#5{%
  \reset@font\fontsize{#1}{#2pt}%
  \fontfamily{#3}\fontseries{#4}\fontshape{#5}%
  \selectfont}%
\fi\endgroup%
\begin{picture}(3941,1789)(508,-499)
\end{picture}%

%% file: Sigma.pstex_t
\begin{picture}(0,0)%
\includegraphics{Sigma.pstex}%
\end{picture}%
\setlength{\unitlength}{4144sp}%
\begingroup\makeatletter\ifx\SetFigFont\undefined%
\gdef\SetFigFont#1#2#3#4#5{%
  \reset@font\fontsize{#1}{#2pt}%
  \fontfamily{#3}\fontseries{#4}\fontshape{#5}%
  \selectfont}%
\fi\endgroup%
\begin{picture}(5002,2840)(1172,-2467)
\put(5356,-1366){\makebox(0,0)[lb]{\smash{{\SetFigFont{12}{14.4}{\rmdefault}{\mddefault}{\updefault}{\color[rgb]{0,0,0}$\Gamma $}%
}}}}
\put(5018,-248){\makebox(0,0)[lb]{\smash{{\SetFigFont{12}{14.4}{\rmdefault}{\mddefault}{\updefault}{\color[rgb]{0,0,0}$Z_G$}%
}}}}
\end{picture}%

%% file: graphs7.bbl
\begin{thebibliography}{}

\end{thebibliography}


\begin{thebibliography}{99}

\bibitem[Ba12]{Ba12} Ram Band.
The Nodal Count $\{0, 1, 2, 3,...\} $  Implies The Graph is a Tree.
{\it Phil. Trans. of the Royal Society }
{\bf A372}:20120504  (2014).

\bibitem[BB13]{BB13}Ram Band \&  Gregory Berkolaiko.
Universality of the momentum band density of periodic networks.
{\it Phys. Rev. Lett.} {\bf 111}:130404 (2013).

\bibitem[BG00]{BG00}Felipe Barra \&  Pierre  Gaspard,
On the Level Spacing Distribution in Quantum Graphs,
{\it Journal of Statistical Physics} {\bf 101}:283--319  (2000).

\bibitem[BG01]{BG01}Felipe Barra \&  Pierre  Gaspard,
Classical dynamics on graphs,
{\it Physical Review E} {\bf 63}:066215 (2001).

\bibitem[BK13]{BK13}
{Gregory Berkolaiko \&  Peter Kuchment},
     {Introduction to quantum graphs},
     {\it Mathematical Surveys and Monographs (AMS)},
     {\bf 186} (2013).




\bibitem[BKW04]{BKW04}
Gregory Berkolaiko, Jonathan   Keating \&  Brian Winn,
No Quantum Ergodicity for Star Graphs.
{\em  Commun. Math. Phys.} {\bf 250}:259--285 (2004).

\bibitem[BW08]{BW08}Gregory Berkolaiko \&  Brian Winn,
Relationship between scattering matrix and spectrum of quantum graphs.
{\it Transactions of the AMS} {\bf 362}:6261--6277 (2010).

\bibitem[Catt97]{Catt97}
Carla Cattaneo.
The Spectrum of the Continuous Laplacian on a Graph.
{\it Monatshefte f\"ur Mathematik} {\bf 124}:215--235 (1997).


\bibitem[CdV85]{CV}
Yves Colin de Verdi\`ere,
\newblock  Ergodicit\'e et fonctions propres du laplacien.
\newblock {\em  Commun. Math. Phys.}  {\bf 102}:497--502, (1985).



\bibitem[Fr05]{Fr05}
{Leonid Friedlander},
    {Genericity of simple eigenvalues for a metric graph},
   {\it Israel J. Math.},
    {\bf 146}:149--156 (2005).

 \bibitem[GKF10]{GKF10}  Sven  Gnutzman, Jon P. Keating \& Fabien  Piotet.
  Eigenfunction
 statistics on quantum graphs,
 {\it Ann. Phys.} {\bf  325}:2595-2640 (2010). 
    


\bibitem[JSS13]{JSS13} Dmitry Jakobson, Yuri Safarov    \& Alexander
  Strohmaier,
The semi-classical theory of discontinuous systems and ray-splitting
billiards (with an Appendix of Yves Colin de Verdi\`ere). 
{\it American Journal of Maths (to appear)} and
{\it ArXiv} 1301.6783v1 (2013).


\bibitem[KMW02]{KMW02} 
Jon P. Keating, Jens Marklof \&  Brian Winn.
Value Distribution of the Eigenfunctions
and Spectral Determinants of Quantum Star Graphs. 
{\it Commun. Math. Phys.} {\bf 241}:421--452 (2003). 

\bibitem[Ni84]{Ni84}
{Serge Nicaise,}
     {Some results on spectral theory over networks, applied to
              nerve impulse transmission},
  {Orthogonal polynomials and applications ({B}ar-le-{D}uc,
              1984)},
    {\it Lecture Notes in Math. (Springer)},
    {\bf 1171}:{532--541} (1985).
 

\bibitem[Ni87]{Ni87}
{Serge Nicaise},
    {Approche spectrale des probl\`emes de diffusion sur les
              r\'eseaux},
 {S\'eminaire de {T}h\'eorie du {P}otentiel, {P}aris, {N}o.\ 8},
 {\it Lecture Notes in Math. (Springer)},
     {\bf 1235}:{120--140} (1987).

\bibitem[SK03]{SK03}Holger Schanz \&  Tsampikos Kottos.
Scars on Quantum Networks Ignore the Lyapunov Exponent.
{\it  Phys. Rev. Lett.} {\bf 90}:234101 (2003).

 \bibitem[Shn74]{Shn74}
Alexander ~I. Shnirelman,
\newblock Ergodic properties of eigenfunctions.
\newblock {\it  Uspehi Mat. Nauk}  {\bf 29}:181--182, (1974).

\bibitem[Shn93]{Shn93} Alexander ~I. Shnirelman,
\newblock On the asymptotic properties of eigenfunctions in the
regions of chaotic motion.
\newblock In V. Lazutkin {\em KAM theory and semiclassical
approximations to eigenfunctions.} {\it Ergebnisse der Mathematik und
ihrer Grenzgebiete (3), 24. Springer-Verlag, Berlin,} (1993].


\bibitem[Zel87]{Zelditch}
Steve Zelditch,
\newblock { Uniform distribution of eigenfunctions on compact
 hyperbolic surfaces.}
\newblock  {\it Duke Math. J.} {\bf 55}:919--941 (1987).



\end{thebibliography}
